%% file: arXivVersion.tex
\documentclass[journal]{IEEEtran}
\usepackage{cite}
\usepackage{amsmath,amssymb,amsfonts}
\usepackage{amsthm}
\usepackage{graphicx,color}
\usepackage{textcomp}
\usepackage{todonotes}
\usepackage{url}
\usepackage{algorithm}
\usepackage{algpseudocode}
\usepackage[nice]{nicefrac}
\usepackage{orcidlink}    
\usepackage{bm,bbm,mathtools}
\usepackage[caption=false,font=footnotesize]{subfig}
\usepackage{tabularx}
\usepackage{multirow}
\usepackage[nolist]{acronym}
\usepackage{balance}
\usepackage{comment}

\newtheorem{theorem}{Theorem}
\newtheorem{lemma}[theorem]{Lemma}

\newcommand{\qmax}{Q_{\mathrm{max}}}
\newcommand{\nrb}{N_{\mathrm{RB}}}
\newcommand{\nre}{N_{\mathrm{RE}}}
\newcommand{\dserv}{\tau_{\mathrm{s}}}
\newcommand{\dwait}{\tau_{\mathrm{w}}}

\newcommand{\dtot}{\tau_\Sigma}
\newcommand{\dvp}{\mathcal{D}}
\newcommand{\snr}{\gamma}

\newcommand{\ddecod}{{\tau_\mathrm{d}}}
\newcommand{\dfeedb}{{\tau_\mathrm{f}}}
\definecolor{iG}{rgb}{0.0, 0.7, 0.0}

\def\BibTeX{{\rm B\kern-.05em{\sc i\kern-.025em b}\kern-.08em
    T\kern-.1667em\lower.7ex\hbox{E}\kern-.125emX}}
\AtBeginDocument{\definecolor{ojcolor}{cmyk}{0.93,0.59,0.15,0.02}}

\input{acronyms}

\begin{document}
\title{{\footnotesize
\begin{tabular}{c}
This work has been submitted to the IEEE for possible publication.\\[-0.4ex]
Copyright may be transferred without notice, after which this version may no longer be accessible.
\end{tabular}
}\\[0.6ex]
Delay Violation Probability Modeling for 5G Systems with HARQ Operation}

\author{Sangwon Seo$^{\star}$\IEEEauthorrefmark{2}\orcidlink{0000-0002-9181-9454},
Vishnu N Moothedath$^{\star}$\IEEEauthorrefmark{2}\orcidlink{0000-0002-2739-5060},\\
Niloofar Mehrnia\IEEEauthorrefmark{2}\orcidlink{0000-0002-5475-2238},
Neda Petreska\IEEEauthorrefmark{3}\orcidlink{0000-0002-2743-5800},
Bernhard Kloiber\IEEEauthorrefmark{3},
James Gross\IEEEauthorrefmark{2}\orcidlink{0000-0001-6682-6559}\\
\IEEEauthorrefmark{2}{Department of Information Science and Engineering, KTH Royal Institute of Technology, Stockholm, Sweden}
\IEEEauthorrefmark{3}{Siemens AG, Munich, Germany}
\thanks{$^{\star}$Sangwon Seo and Vishnu N Moothedath are the co-first authors.}
\thanks{Corresponding author: Sangwon Seo, (e-mail: sangwon@kth.se)}}

\maketitle

\begin{abstract}
Meeting the growing demand for quality-of-service (QoS) guarantees in 5G networks requires an accurate characterization of delay performance, commonly captured by the delay violation probability (DVP) at a specified delay target. Although hybrid automatic repeat request (HARQ) is a fundamental reliability mechanism in wireless systems and is central to supporting QoS, many existing approaches to DVP prediction for HARQ remain overly simplified. In particular, they omit important delay components and adopt assumptions that do not reflect the operation of HARQ in slot-based systems such as 5G. Consequently, these models can substantially underestimate the DVP, especially under stringent latency requirements, where the contribution of the neglected components becomes critical. To address this gap, we develop a tractable DVP characterization for 5G HARQ that accounts for queueing, transmission, decoding, and feedback delay, as well as the contribution of \ac{CS} transmissions to the overall delay, under practical timing assumptions consistent with 3GPP operation. Moreover, we incorporate parallel packet transmissions that proceed without waiting for earlier packets to succeed, an essential HARQ behavior frequently overlooked in prior work. Using tools from queueing theory and Markov analysis, we then derive upper bounds on the DVP and validate them against ns-3 5G-LENA simulations.
\end{abstract}

\bstctlcite{IEEEexample:BSTcontrol}

\section{Introduction}\label{sec:intro}
The advent of 5G networks has fundamentally transformed wireless communication by supporting heterogeneous service categories such as \ac{URLLC}, \ac{eMBB}, and \ac{mMTC}~\cite{3gpp2017study,popovski20185G,3gpp.22.104,3gpp.22.261}. Among these, \ac{URLLC} places the most stringent \ac{QoS} requirements on latency and reliability and is therefore a key enabler for real-time applications, including autonomous driving, virtual reality, and Industry~4.0~\cite{sisinni2018industrial}. Such applications typically involve short packets and moderate throughput~\cite{durisi2016URLLC}, yet they demand end-to-end delays on the order of milliseconds and \ac{PER} as low as $10^{-3}$~\cite{popovski20226g} down to $10^{-5}$~\cite{3gpp.38.913} for communication among devices such as sensors, actuators, and controllers.
A widely used metric in reliability-latency trade-off is the \ac{DVP}, defined as the probability that a packet's end-to-end delay exceeds a target threshold. The \ac{DVP} is shaped by the cumulative impact of multiple delay contributions, including queueing, transmission, decoding/processing, and \ac{HARQ} feedback-related timing. Moreover, while \ac{CS} is not itself a delay component, the time spent transmitting and waiting for control signals (e.g., scheduling-related signaling) can directly increase packet delay.

A straightforward approach to meet stringent reliability targets is to strengthen channel coding to drive the \ac{PER} as low as possible. Under tight latency constraints, however, this strategy is often inefficient and not always feasible. In particular, more powerful codes and more complex decoding can incur non-negligible processing/decoding delays, which may offset (or even outweigh) the gains from a lower error rate and thus degrade the overall \ac{QoS}~\cite{celebi2021latency}. Moreover, for given channel conditions and a fixed latency budget, minimizing the \ac{PER} alone is generally suboptimal~\cite{Peng2011ReliabilityPHY}. A more effective operating point is a moderate error rate in conjunction with retransmission mechanisms. In this context, \ac{ARQ}~\cite{ArqCommEngDeskRef} and \ac{HARQ}~\cite{Frenger2001HarqHSDPA,DAHLMAN2014299} are widely adopted schemes that enhance reliability via feedback and selective retransmissions.

\ac{ARQ}~\cite{ArqCommEngDeskRef} enhances communication reliability by retransmitting packets that receive \acp{NACK}. In its most basic implementation, \ac{ARQ} operates as a sequential \ac{FIFO} service. However, this approach can lead to inefficient use of transmission resources, as the transmitter must often wait for feedback before sending subsequent packets. \ac{HARQ}~\cite{Frenger2001HarqHSDPA,DAHLMAN2014299} extends the capabilities of \ac{ARQ} in two significant ways. First, through \ac{IR}, each retransmission delivers additional coded bits, enabling the receiver to accumulate information from multiple transmission attempts and thereby incrementally increasing the probability of successful decoding. Second, modern systems typically support several parallel \ac{HARQ} processes, allowing new packets to be transmitted even while previous ones are still awaiting acknowledgment. In this pipelined framework, each packet is associated with a unique \ac{HARQ} process identifier, thereby permitting out-of-order completion of reception. This design reduces idle time caused by feedback delays and optimizes resource utilization. These features make \ac{HARQ} particularly advantageous for scenarios involving short packets and low-latency requirements, where limited time diversity amplifies the need for feedback-driven reliability improvements~\cite{Ostman2018Harq}.

While \ac{HARQ} is widely recognized as a key mechanism for meeting stringent reliability targets in short-packet, low-latency regimes, its impact on delay performance is not captured adequately by much of the existing \ac{DVP} analysis. Specifically, several previous studies on HARQ-related \ac{DVP}~\cite{Sahin2014Harq,Sahin2015Harq,Sahin2019Harq} employ simplified single-process models, such as \ac{FIFO} abstractions, which fail to represent key aspects of practical 5G \ac{HARQ} systems, most notably, the use of parallel \ac{HARQ} processes that enable pipelined transmissions and allow packets to complete out of order. Furthermore, these models frequently omit delay components that become critical under strict latency constraints in slot-based systems, including decoding and processing times, feedback delays, and \ac{CS}-caused delay. Consequently, such simplifications can lead to overly optimistic \ac{DVP} estimations when these factors constitute a significant portion of the total end-to-end delay. To bridge this gap, we propose a tractable \ac{DVP} framework for 5G \ac{HARQ} that explicitly models parallel-process operation and incorporates these sources of delay in practice.

\subsection{Related Work}
A first line of work established foundational tools for analyzing wireless delay by characterizing queueing behavior and deriving delay bounds via queueing theory, large deviations, and network-calculus-based methods~\cite{Telatar1995,bisnik2006queuing,chang1995EB,wu2003effectiveCap,hassan2004markov,fidler2006wlc15,jiang2008SNC,Ciucu2010,Zubaidi2016,devassy2019reliable,yeh2012fundamental,petreska2019bound}. For instance, \cite{Telatar1995} studies the trade-off between error probability and delay in multi-access \ac{AWGN} systems using queueing-theoretic arguments. Additionally, \cite{bisnik2006queuing} develops analytical end-to-end delay models for wireless networks represented as G/G/1 queues. 
In parallel, effective bandwidth \cite{chang1995EB} and effective capacity \cite{wu2003effectiveCap} frameworks were used to connect service variability to delay and error performance \cite{hassan2004markov,fidler2006wlc15}, while stochastic network calculus \cite{jiang2008SNC,Ciucu2010}, (min,$\times$) algebra~\cite{Zubaidi2016}, and related large-deviation techniques \cite{yeh2012fundamental,petreska2019bound,wu2003effectiveCap,fidler2006wlc15} enabled tractable delay bounds and, in some cases, delay distributions. However, they typically abstract away retransmission mechanisms for improved reliability, including \ac{HARQ}, and their timing dynamics in \ac{DVP} analysis in modern systems.

To address this missing reliability mechanism, a second body of work explicitly incorporates \ac{ARQ}/\ac{HARQ} into delay or service-process modeling, often by embedding retransmissions into effective-capacity or state-transition formulations. For instance, \cite{larsson2016HARQ} analyzes \ac{HARQ} via effective capacity, and~\cite{akin2015backlog} develops a state-transition model for effective capacity where packet errors are driven by outage events defined through Shannon capacity~\cite{Shannon}. These works importantly move beyond retransmission-agnostic delay models by making feedback-based reliability part of the analysis. Nonetheless, they largely rely on asymptotic information-theoretic arguments (e.g., outage/ergodic capacity) that presume large packet sizes \cite{devassy2019reliable} and tend to prioritize throughput-oriented metrics over stringent tail-delay guarantees. 
As a result, they can be inaccurate for short-packet, \ac{DVP}-critical operation, where Shannon-based abstractions are known to overestimate performance \cite{durisi2016URLLC}, as also highlighted by the \ac{FBL} delay analysis~\cite{schiessl2015delay}.

Motivated by this short-packet mismatch, subsequent studies adopt \ac{FBL} models to better capture the reliability-latency interplay under short packet sizes. 
Building on the \ac{FBL} framework~\cite{polyanskiy2010FBL,polyanskiy2011feedback,Wei2013FBLblockfading}, \cite{schiessl2018delay} extends their earlier delay analysis to the FBL regime. 
Moreover, \cite{devassy2014finite} integrates \ac{FBL} capacity into \ac{HARQ}-inspired models (with a focus on \ac{ARQ}) to estimate throughput and delay. Importantly for \ac{QoS}, \cite{devassy2018delay,devassy2019reliable} analyze the \ac{DVP} for \ac{ARQ} (treating \ac{HARQ} as a special case) and show that similar average delay can correspond to markedly different \ac{DVP}, underscoring that mean-delay metrics are insufficient for stringent reliability requirements. 
Related \ac{FBL}-based queueing analyses for \ac{HARQ} are provided by \cite{Sahin2014Harq,Sahin2015Harq,Sahin2019Harq}. While these works provide a performance characterization of the queueing dynamics, they simplify system timing by taking only waiting delays into consideration and assuming negligible decoding delay and instantaneous feedback.
Under such assumptions, \ac{HARQ} effectively collapses to a sequential \ac{FIFO}-like service that cannot represent the parallel, pipelined \ac{HARQ} operation of a real slot-based 5G.

A more system-oriented direction attempts to reintroduce practical protocol effects by accounting for additional delay components and feedback imperfections under \ac{FBL} operation. For example, \cite{FixedTx} incorporates multiple delay contributions and feedback failures, thereby moving closer to implementation-relevant timing than models with instantaneous acknowledgments. However, this refinement is achieved for \ac{ARQ} and retains a sequential single-server/\ac{FIFO} abstraction, leaving out a defining characteristic of practical \ac{HARQ}, namely multiple parallel \ac{HARQ} processes that allow new packets to be transmitted while earlier packets await feedback, thereby enabling out-of-order packet completion. Moreover, existing models typically do not fully capture the compounded impact of different delay components in slot-based 5G, where non-negligible \acp{RTT} invalidate assumptions such as transmission-dominated \acp{RTT} (e.g., \cite{Sahin2019Harq}).

\subsection{Contributions}
In this paper, we propose a tractable \ac{DVP} characterization for 5G \ac{HARQ} that explicitly models parallel \ac{HARQ} processes and incorporates queueing, transmission, decoding/processing, feedback timing, and \ac{CS} transmissions under practical 3GPP-consistent timing.
The main contributions of this work are summarized as follows:
\begin{enumerate}
\item We propose a \ac{DVP} characterization framework for 5G \ac{HARQ} that models the parallel \ac{HARQ} processes using a multi-server queue and captures various delay components accounting for practical slot-based timing effects.
\item We derive an analytical upper bound on the \ac{DVP} for 5G \ac{HARQ}, in the presence of decoding delay, feedback delay, and periodic \ac{CS}.
A \texttt{MATLAB}-based validation confirms that the derived result provides a tight upper bound on the simulated \ac{DVP}.
\item We demonstrate that the \ac{DVP} bound obtained by the proposed model remains closely aligned with empirical \ac{DVP} from the \texttt{ns-3} 5G-LENA simulator across a broad range of scenarios, including different delay targets, arrival rates, \ac{SNR}, and \ac{CSP}, while outperforming existing methods for estimating the \ac{DVP} bound.
\end{enumerate}

The remainder of this work is organized as follows: In Section~\ref{sec:model}, we introduce the system model and problem statement. 
In Section~\ref{sec:harq}, we provide the proposed framework for deriving the \ac{DVP}, considering the \ac{HARQ} retransmission scheme and the delay components corresponding to this scheme. 
Finally, in Section~\ref{sec:simulations}, we show the numerical evaluation and conclude in Section~\ref{sec:conclusion}.
Important notations are listed in TABLE~\ref{Tab:notations}.

\begin{table}
\centering
\setlength{\belowcaptionskip}{6pt}
\caption{List of important notations.}
\label{Tab:notations}
\setlength{\tabcolsep}{3pt}
\begin{tabular}{|p{35pt}|p{195pt}|}
\hline
{Notation} & {Description} \\
\hline
$n$       & Packet length (bits) \\
$T$       & Slot duration (s) \\
$f$       & Frequency of random arrivals, with $f<1$ \\
$\qmax$   & Maximum queue size \\
$d$       & Delay target (s) \\
$\dvp(d)$ & DVP for the delay target $d$ \\
$\nrb$    & Number of RBs \\
$\nre$    & Number of REs \\
$\xi$     & Control signal periodicity (CSP), in slots \\
$\dtot$   & Total delay, in slots \\
$\dserv$  & Service delay, in slots \\
$\dwait$  & Waiting delay, in slots \\
$\dfeedb$ & Feedback delay, in slots \\
$\ddecod$ & Decoding delay, in slots \\
$m$       & Transmission index \\
$p_m$     & PER of the $m^{\text{th}}$ HARQ attempt \\
$M$       & Maximum transmission attempts \\
$M_{d/T}$ & Maximum transmission attempts possible without violating the delay target \\
$\eta$    & Spectral efficiency \\
$\snr$    & Average SNR \\
\hline
\end{tabular}
\end{table}

\section{System Model}\label{sec:model}
\begin{figure*}[th]
\centering
\begin{minipage}[t]{.5\textwidth}
  \centering
  \includegraphics[width=\textwidth]{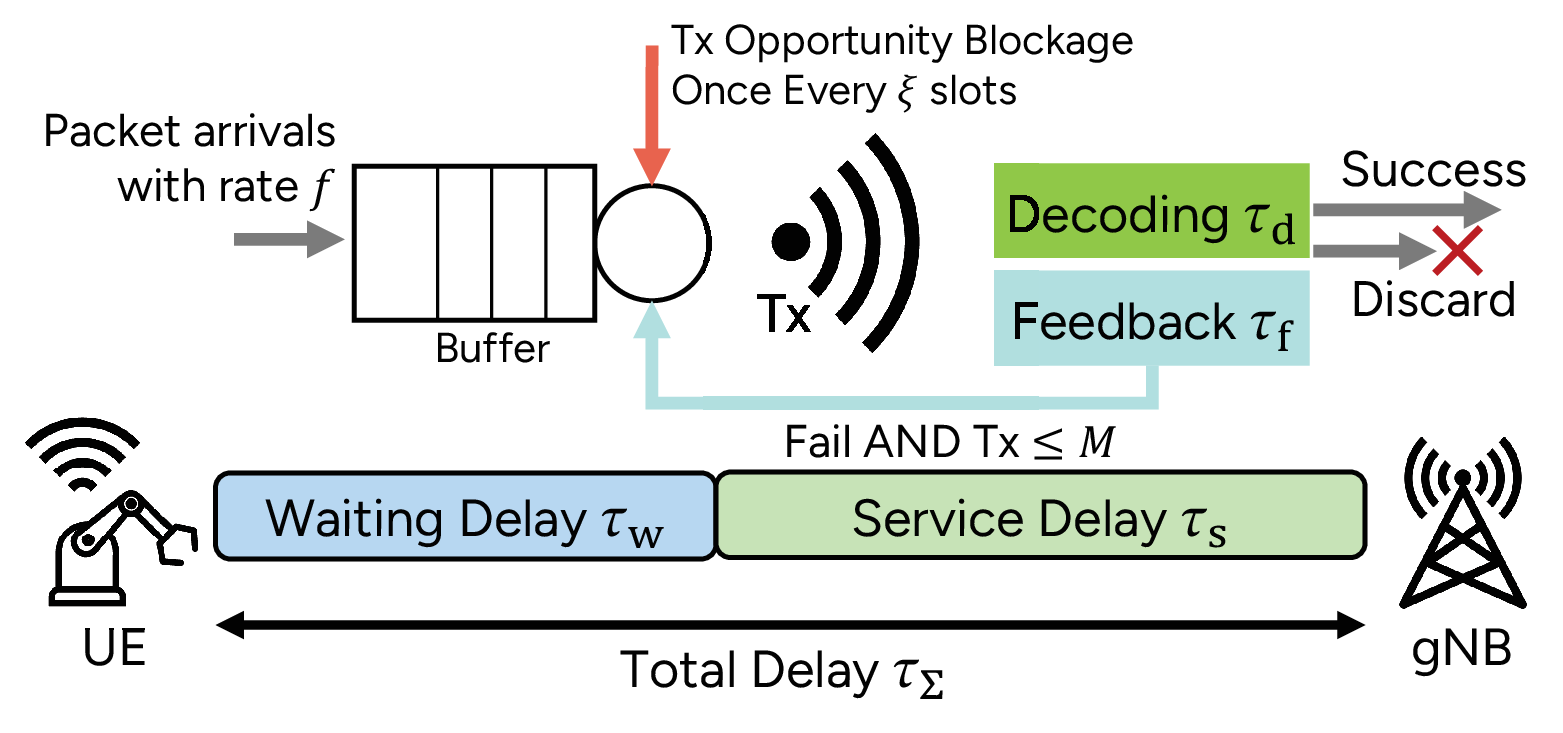}
  \caption{System model, showing the retransmission process. Different delay components are shown where the packets experience them.
}
  \label{fig:BasicModel}
\end{minipage}%
\hspace{0.03\textwidth}%
\begin{minipage}[t]{0.46\textwidth}
  \centering
  \includegraphics[width=\textwidth]{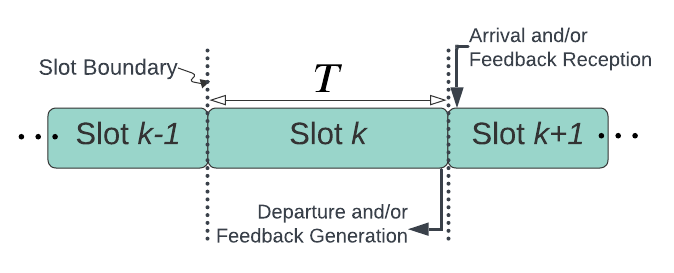}
  \caption{Timing diagram showing the order and positions of different slot-based arrival and departure events with respect to the corresponding slot boundary.}
  \label{fig:timingdiagram}
\end{minipage}
\end{figure*}

We consider a slot-based 5G-\ac{NR} communication link with \ac{HARQ} retransmissions, as shown in Figure~\ref{fig:BasicModel}. Time is divided into slots of fixed duration $T$. Packets of size $n$ bits arrive randomly at the \ac{UE}, are buffered in a finite queue of size $\qmax$\footnote{The analysis applies to both uplink and downlink; we focus on the uplink for clarity.}, and are transmitted to a dedicated \ac{gNB} over a 5G-\ac{NR} wireless link. In each \emph{available} slot, the \ac{UE} is scheduled a fixed amount of radio resources (a fixed number of \acp{RB}) and can transmit at most one packet (either a new packet or a retransmission) using a configured \ac{MCS}. \ac{HARQ} operates with multiple parallel processes, allowing pipelined transmissions so that the \ac{UE} need not wait for the \ac{ACK}/\ac{NACK} of earlier packets before sending new ones. Receiver processing/decoding and feedback timing are explicitly modeled via non-zero decoding and feedback delays.

The queue size is measured at the slot boundary. In each slot $k$ illustrated in Figure~\ref{fig:timingdiagram}, a new packet may arrive at the queue, and any \ac{ACK}/\ac{NACK} whose feedback delay has elapsed is received at the \ac{UE}, right after the slot boundary. 
If slot $k$ is available for data and the \ac{UE} schedules a transmission, it occupies the entire time domain resources of the slot for the packet transmission. 
After the receiver processing (subject to the decoding delay model), the \ac{gNB} triggers feedback, right before the corresponding slot boundary.
Slot indices at the \ac{UE} and \ac{gNB} are assumed synchronized via timing advance (TA) procedure, such that a \ac{UE} transmission in slot $k$ is received within slot $k$ at the \ac{gNB}~\cite{3gpp.38.321}.

Some slots are periodically unavailable for data due to \ac{CS} transmissions. Let $\xi$ denote the \ac{CSP} shown in Figure~\ref{fig:BasicModel}: one out of every $\xi$ slots is reserved for \ac{CS} transmission and thus blocks the corresponding data transmission opportunity. We assume $\xi$ is much larger than the delay target, consistent with typical 5G configurations~\cite{3gpp.38.211}.

Packet arrivals are modeled as an \ac{i.i.d.}\ Bernoulli process, where, in each slot, a new packet arrives with probability $f$. Arriving packets join a queue of capacity $\qmax$ packets; if the queue is full upon arrival, the packet is dropped (overflow).
Upon arrival, the packet waits in the queue in \ac{FIFO} order until it becomes head-of-line and a data slot becomes available for transmission. This waiting time accounts for both queue backlog and periodic \ac{CS} blockage. When scheduled for its first transmission, the packet is dequeued and assigned to a \ac{HARQ} process (identified by a \ac{HARQ} ID). From this point on, it is stored outside the queue (within the \ac{HARQ} process state) until it is either successfully delivered or discarded after exhausting the allowed number of attempts.

After each transmission attempt, the \ac{gNB} spends $\ddecod$ slots for receiver processing/decoding before the decoding outcome is available. Once available, the \ac{gNB} generates feedback \ac{ACK} if the packet is decoded successfully, and \ac{NACK} if decoding fails.
The feedback is then received at the \ac{UE} after an additional $\dfeedb$ slots. Hence, each attempt incurs an explicit decoding-and-feedback timing of $\ddecod+\dfeedb$ slots before the \ac{UE} can react. If an \ac{ACK} is received, the packet departs the system successfully. If a \ac{NACK} is received, the packet becomes pending for retransmission. A packet is allowed at most $M$ (re)transmission attempts: if $M<\infty$ the scheme is \emph{truncated} and the packet is discarded after $M$ failed attempts; if $M=\infty$ the scheme is \emph{persistent}. We assume that \acp{NACK} for failed packets are always delivered successfully in the \ac{DL}~\cite{ErrorFreeHARQFeed}. 

Parallel \ac{HARQ} enables pipelining so that while one packet is waiting for decoding/feedback, other packets (possibly newer ones) may be transmitted using different \ac{HARQ} processes. We assume at most one transmission per (data) slot. We additionally give priority to schedule the retransmission if one or more \ac{HARQ} processes have a pending retransmission (i.e., a \ac{NACK} has been received and another attempt is allowed). Otherwise, if no retransmission is pending and the queue is non-empty, the head-of-line packet is scheduled for a first transmission.

As highlighted in Figure~\ref{fig:BasicModel}, we decompose the total delay (in slots) as
\begin{equation}
    \dtot = \dwait + \dserv,
\end{equation}
where $\dwait$ is the waiting delay from arrival until the first transmission, and $\dserv$ is the service delay from the first transmission until final success (\ac{ACK}) or discard (after $M$ failures), including decoding/processing time, feedback timing, \ac{CS} blockage, and any retransmissions.
For a delay target $d$ (seconds), the \ac{DVP} is defined as
\begin{align}
    \dvp(d) &= \mathbb{P}\!\left(\dtot > d/T\right).
    \label{eq:SecModel_dvpDefinition}
\end{align}

Encoding delay is typically ignored since it occurs once per packet and can be overlapped with queueing. If deemed necessary, it can be incorporated by reducing the effective delay budget. Additionally, since the system is slotted, we skip modeling the in-slot transmission time explicitly. However, it can be integrated into the receiver-side processing/decoding component.

5G-\ac{NR} uses \ac{OFDM} and allocates resources in quanta of \acp{RB}. Let $\nrb$ denote the number of \acp{RB} allocated to the \ac{UE} per transmission attempt. Each \ac{RB} contains 12 subcarriers with subcarrier spacing $15\times 2^\nu$ kHz, where $\nu$ denotes the numerology~\cite{3gpp.38.211}. We denote by $\nre$ the number of resource elements across the allocated \acp{RB} per \ac{OFDM} symbol, i.e., $\nre = 12\nrb$.

The system assumes \ac{HARQ}-\ac{IR}~\cite{DAHLMAN2014299}, where the packet error rate depends on the attempt index $m$:
\begin{equation}
    \bm{p} = [p_1,p_2,\dots,p_M], \qquad p_m \ge p_{m+1},
\end{equation}
where $p_m$ is the \ac{PER} of the $m^{\text{th}}$ (re)transmission attempt. We model
\begin{equation}
    p_m=\Phi(\gamma,\eta,m,\nre),
    \label{eq:secModel_perHARQ_avgSNR}
\end{equation}
where $\gamma$ and $\eta$ denote the average \ac{SNR} and spectral efficiency, respectively, and $\Phi(\cdot)$ is a generic \ac{PER} abstraction obtained from link-level simulation, measurements, or analytical approximations~\cite{Sangwon2025,polyanskiy2010FBL,polyanskiy2011feedback,Vishnu2025}.

Given $(f,\qmax,\xi,\ddecod,\dfeedb,M)$ and the per-attempt error probabilities in \eqref{eq:secModel_perHARQ_avgSNR}, our objective is to characterize the \ac{DVP} in \eqref{eq:SecModel_dvpDefinition} as a function of the system parameters under practical slot-based 5G-\ac{NR} with \ac{HARQ} timing with parallel processes.

\begin{table*}
\centering
\setlength{\belowcaptionskip}{6pt}
\caption{Non-zero transition probabilities of $P$ from a given pair of queue size $q$ and HARQ tranmission index $m$.}
\begin{tabular}{|l|l|l|l|l|}
\hline
\multicolumn{1}{|c|}{State} &
\multicolumn{1}{c|}{Next state} &
\multicolumn{1}{c|}{Range $(q)$} &
\multicolumn{1}{c|}{Range $(m)$} &
\multicolumn{1}{c|}{Probability} \\\hline
$(0,1)$  & $(0,1)$ & - & - &
$1-(1-I(k_\xi)) f - I(k_\xi) f p_1$
\\
$(0,1)$  & $(1,1)$ & - & - &
$(1-I(k_\xi)) f$
\\
$(0,1)$  & $(1,2)$ & - & - &
$I(k_\xi) f p_1$
\\\hline
$(q,1)$ & $(q,1)$ & $[1,Q_{\max})$ & - &  $(1-I(k_\xi)) f' + I(k_\xi)\big(f p_1'\big)$
\\
$(q,m)$  & $(q,m)$     & $[1,Q_{\max})$ & $[2, M)$  &
$(1-I(k_\xi)) f'$
\\
$(q,m)$  & $(q+1,m)$   & $[1,Q_{\max})$ & $[1, M)$  &
$(1-I(k_\xi)) f$
\\
$(q,m)$  & $(q,m+1)$   & $[1,Q_{\max})$ & $[1,M)$  &
$I(k_\xi) f' p_m$
\\
$(q,m)$  & $(q+1,m+1)$ & $[1,Q_{\max})$ & $[1,M)$ &
$I(k_\xi) f p_m$
\\
$(q,m)$  & $(q,1)$     & $[1,Q_{\max})$ & $[2,M)$ &
$I(k_\xi) f p_m'$
\\
$(q,m)$  & $(q-1,1)$   & $[1,Q_{\max})$ & $[1,M)$ &
$I(k_\xi) f' p_m'$
\\
\hline
$(q,M)$  & $(q,1)$     & $[1,Q_{\max})$ & - &
$I(k_\xi) f$
\\
$(q,M)$  & $(q-1,1)$   & $[1,Q_{\max})$ & - &
$I(k_\xi) f'$
\\
$(q,M)$  & $(q,M)$     & $[1,Q_{\max})$ & - &
$(1-I(k_\xi)) f'$
\\
$(q,M)$  & $(q+1,M)$   & $[1,Q_{\max})$ & - &
$(1-I(k_\xi)) f$
\\
\hline
$(Q_{\max},m)$ & $(Q_{\max},m)$ & - & $[1,M]$ &
$1-I(k_\xi) $
\\
$(Q_{\max},m)$ & $(Q_{\max},m+1)$ & - & $[1,M)$ &
$I(k_\xi)\,p_m$
\\
$(Q_{\max},m)$ & $(Q_{\max}-1,1)$ & - & $[1,M)$ &
$I(k_\xi)\,p_m'$
\\
$(Q_{\max},M)$ & $(Q_{\max}-1,1)$ & - & - &
$I(k_\xi) $
\\\hline
\end{tabular}
\label{tab:new_qmk_onoff}
\end{table*}

\section{HARQ-based DVP Modeling}\label{sec:harq}
This section develops a tractable framework to upper-bound the \ac{DVP} of slot-based \ac{HARQ} under the system model of Section~\ref{sec:model}. The key idea is to first characterize the \emph{stationary backlog} seen by an arriving packet via a Markov-chain description of the slot evolution, translate that backlog into a \emph{waiting-delay} distribution, and combine it with a \emph{service-delay} violation characterization that explicitly accounts for decoding and feedback timing. Throughout, we measure delays in \emph{slots} and use the shorthand $\bar d \triangleq d/T$ for a delay target of $d$ seconds.

\subsection{Slot-level Markov model for backlog}\label{subsec:harq_markov}

We model the system evolution at slot boundaries by a \ac{DTMC} whose state captures (a) the backlog in the \ac{UE} buffer and (b) the position within the \ac{CSP}. Recall that one out of every $\xi$ slots is reserved for \ac{CS} and therefore does not provide a data transmission opportunity.

Let $q\in\{0,1,\dots,Q_{\max}\}$ denote the number of packets in the \ac{UE} buffer (i.e., in the queue), immediately after a slot boundary. Let also $m\in\{1,2,\dots,M\}$ denote the \ac{HARQ} transmission-attempt index associated with the packet served by the scheduler when a data transmission occurs. Finally, let
$k_\xi\in\{0,1,\dots,\xi-1\}$ denote the \ac{CSP} phase, which is relative slot position within a \ac{CSP} cycle. We use the indicator
\begin{equation}
    I(k_\xi)=
    \begin{cases}
        0, & k_\xi = 0,\\
        1, & k_\xi \in \{1,\dots,\xi-1\},
    \end{cases}
\end{equation}
where $I(k_\xi)=0$ indicates a blocked slot reserved for \ac{CS} and $I(k_\xi)=1$ indicates an available data slot. The \ac{CSP} phase evolves deterministically as
\begin{equation}
    k_\xi \leftarrow (k_\xi+1)\bmod \xi .
\end{equation}
Here, $\bmod \,\, \xi$ denotes the modulo (remainder) operation. $(k_\xi+1)\bmod \xi$ is the unique integer in $\{0,1,\ldots,\xi-1\}$ equal to the remainder when $k_\xi+1$ is divided by $\xi$.

The \ac{DTMC} has $(Q_{\max}+1)M\xi$ states. For implementation convenience, we map $(q,m,k_\xi)$ to a single state index
\begin{equation}
    s = k_\xi (Q_{\max}+1)M + qM + (m-1),
\end{equation}
so that $s\in\{0,1,\dots,(Q_{\max}+1)M\xi-1\}$.
The state transitions depend on three main factors: (i) the Bernoulli arrival process with arrival rate $f$ per slot, (ii) whether the current slot is blocked ($I(k_\xi)=0$) or available ($I(k_\xi)=1$), and (iii) the decoding outcome of the transmitted packet when a transmission occurs, governed by the attempt-dependent \ac{PER} vector $\bm{p}=[p_1,\dots,p_M]$. Boundary conditions, including empty/full queue and $m=M$, are handled explicitly in the transition rules (see Table~\ref{tab:new_qmk_onoff}). For notational brevity, we define $f'\triangleq 1-f$ and $p_m'\triangleq 1-p_m$.

Let $P\in\mathbb{R}^{(Q_{\max}+1)M\xi \times (Q_{\max}+1)M\xi}$ be the transition matrix. The steady-state distribution of state $S=(q,m,k_\xi)$, denoted by $\tilde{\bm{\pi}}$, satisfies $\tilde{\bm{\pi}}^T = \tilde{\bm{\pi}}^T P$ (equivalently, $P^T\tilde{\bm{\pi}}=\tilde{\bm{\pi}}$).
$\tilde{\bm{\pi}}$ can be computed by finding the eigenvector of $P^T$ corresponding to the unit eigenvalue, which in turn can be computed using standard algorithms or by iterating $P$ until $P^i\!\approx\!P^{i+1}$, for some integer $i$, with the rows converging to the steady-state probabilities.
Because multiple DTMC states share the same queue length $q$, we extract the stationary queue-length distribution $\{\pi_q\}$ by aggregation:
\begin{equation}
    \pi_q
    =
    \sum_{k_\xi=0}^{\xi-1}\sum_{m=1}^{M}
    \tilde{\pi}_{\,k_\xi (Q_{\max}+1)M + qM + (m-1)} .
    \label{eq:pi_q}
\end{equation}

Packets may be removed from the system if they reach the \ac{HARQ} attempt limit $M$ (discard after $M$ failures), and if the queue overflows when $q=Q_{\max}$ and a new arrival occurs. The \ac{DTMC} construction in Table~\ref{tab:new_qmk_onoff} accounts for the relevant boundary behavior. In the operating regimes of interest, i.e., stable queue and sufficiently large $Q_{\max}$, overflow events are typically negligible. Nonetheless, the analysis below is written for finite $Q_{\max}$.

\subsection{Waiting delay}\label{subsec:harq_wait}
An arriving packet to the queue experiences a \emph{waiting delay} $\dwait$ (in slots) until its first transmission. We obtain the \ac{PMF} of $\dwait$ in two steps. First, we compute an underlying conditional waiting-delay \ac{PMF} $f^{\mathrm{u}}_{\dwait}(k\mid q)$ given that the packet sees $q$ packets ahead of it in the transmission order (Algorithm~\ref{Alg:getWaitProbability}). Second, we incorporate \ac{CSP} blockage to obtain $f_{\dwait}(k\mid q)$ and then average over $\pi_q$.

Algorithm~\ref{Alg:getWaitProbability} takes as input the target wait length $k$, the number of packets ahead $q$, the PER vector $\bm{p}$, the current recursion depth (maximum attempts) $M$, and the original maximum $M_0$ used to detect the outermost call. It first checks whether $k$ equals the minimum feasible wait $q$ (Line~1). If $k=q$, every one of the $q$ packets must succeed on its first transmission attempt, which occurs with probability $(1-p_1)^q$, so the function returns this value immediately (Line~2). If $k$ is outside the feasible range, i.e., $k<q$ (too short to clear $q$ packets) or $k>Mq$ (too long even if each packet consumes $M$ attempts), the function returns probability zero (Lines~3--4). If neither base case applies, an accumulator \texttt{prob} is initialized to sum the probability of all feasible configurations that yield total wait $k$ (Line~6). The algorithm then determines the largest feasible number $N$ of packets (among the $q$ ahead) that can consume exactly $M$ attempts while still allowing the total wait to equal $k$ (Line~7). It enumerates the number $n\in\{0,1,\ldots,N\}$ of packets (among the $q$ ahead) that are assumed to consume exactly $M$ attempts at the current recursion level (Line~8). For each such choice of $n$, it computes the combinatorial multiplicity $\binom{q}{n}$, i.e., how many ways $n$ packets can be selected from the $q$ packets ahead to follow the $M$-attempt pattern (Line~9). It computes the probability that a single selected packet reaches $M$-th attempt, which requires failing attempts $1,2,\ldots,M-1$, giving the factor $\prod_{i=1}^{M-1} p_i$ (Line~10). Next, it determines the contribution of the $M$-th attempt for those selected packets by distinguishing two cases: outermost call versus inner recursion (Lines~11--15). At the outermost call ($M=M_0$), the packet leaves the queue after the $M_0$-th attempt regardless of whether it succeeds or is discarded, so the last-attempt factor is set to $1$ (Lines~12--13). In inner recursion levels ($M<M_0$), the reduced subproblem counts only removals that occur by success at $M$-th attempt, hence the last-attempt factor is set to $(1-p_M)$ (Lines~14--15). Using the above factors, it forms the probability that \emph{all} of the $n$ selected packets follow the same $M$-attempt pattern, and raises the single-packet probability to the power $n$ (Line~16). It then defines the residual subproblem obtained after accounting for these $n$ packets: the remaining wait target becomes $k-Mn$, the remaining number of packets becomes $q-n$, and the maximum attempts for the residual packets becomes $M-1$ (Line~17). The function calls itself recursively to compute the probability of this residual subproblem. Finally, it adds the contribution of the current configuration to \texttt{prob} by multiplying (i) the multiplicity $\binom{q}{n}$, (ii) the probability of the selected $n$ packets’ $M$-attempt pattern, and (iii) the probability returned by the residual recursive call (Line~18). After all feasible values of $n$ have been enumerated, it returns the accumulated probability \texttt{prob}, which equals $f^{\mathrm{u}}_{\dwait}(k\mid q)$ (Lines~19--20).

\begin{algorithm}[t]
\caption{\textproc{getWaitProbability} to compute $f^{\mathrm{u}}_{\dwait}(k\!\mid\!q)$.}
\label{Alg:getWaitProbability}
\begin{algorithmic}[1]
\item[]\hspace{-\algorithmicindent}  \textbf{Function} \textproc{getWaitProbability}$(k, q, \bm{p}, M, M_0)$
    \If{$k == q$}
        \State \Return $(1 - {p}_1)^q$ \Comment{$k=q$ is the minimum wait: all $q$ packets succeed at attempt 1.}
    \ElsIf{$k < q$ \textbf{or} $k > M \cdot q$}
        \State \Return $0$ \Comment{Infeasible wait length.}
    \EndIf
    \State $prob \gets 0$
    \State $N \gets \min\!\left(\left\lfloor\frac{k - q}{M - 1}\right\rfloor,\, q\right)$
    \For{$n = 0$ \textbf{to} $N$}
        \State $numSeqs \gets \dbinom{q}{n}$
        \State $seqProb\_failures \gets \prod_{i=1}^{M-1}{p}_{i}$
        \If{$M == M_0$}
            \State $seqProb\_last \gets 1$ \Comment{At the outermost level, success or discard both remove the packet.}
        \Else
            \State $seqProb\_last \gets (1 - {p}_M)$ \Comment{In inner levels, the packet must succeed at $M$-th attempt.}
        \EndIf
        \State $seqProb \gets (seqProb\_failures \cdot seqProb\_last)^n$
        \State $subSeqProb \gets$ \textproc{getWaitProbability}$(k-Mn, q-n, \bm{p}, M-1, M_0)$
        \State $prob \gets prob + numSeqs \cdot seqProb \cdot subSeqProb$
    \EndFor
    \State \Return $prob$
\item[]\hspace{-\algorithmicindent}  \textbf{End Function}
\end{algorithmic}
\end{algorithm}

The \ac{CSP} introduces occasional blocked slots such that, over a waiting interval of $k$ slots, the probability that at least one blocked slot occurs can be upper-bounded by
\begin{equation}
    \beta_k \triangleq \min\left\{1,\frac{k}{\xi}\right\},
\end{equation}
which follows from a union bound and is accurate when $q\ll \xi$.

If no blocked slot is encountered, the waiting delay remains $k$; if at least one blocked slot occurs, the waiting delay increases by (at most) one slot in our bound. Therefore, for each $q$,
\begin{equation}
    f_{\dwait}(k\mid q)
    \leq
    \left(1-\beta_k\right) f_{\dwait}^{\mathrm{u}}(k\mid q)
    +
    \beta_k\, f_{\dwait}^{\mathrm{u}}(k-1\mid q).
    \label{eq:secIR_waitDelay_conditional_xi}
\end{equation}
Finally, averaging over the stationary queue-length distribution yields the unconditional \ac{PMF}
\begin{equation}
    f_{\dwait}(k)=\mathbb{P}(\dwait=k)
    \leq
    \sum_{q=0}^{Q_{\max}} \pi_q\, f_{\dwait}(k\mid q).
    \label{eq:secIR_waitDelay}
\end{equation}

\subsection{Service delay and total \ac{DVP} bound}\label{subsec:harq_service_total}
After the first transmission of the tagged packet, it experiences a \emph{service delay} $\dserv$ that includes the slots used by its (re)transmissions, decoding/processing delay $\ddecod$, and feedback delay $\dfeedb$ between attempts. The packet departs successfully upon receiving an \ac{ACK} (successful decoding) and remains in service upon receiving a \ac{NACK} (decoding failure), up to the attempt cap $M$, after which it is discarded.

Let $\bar d$ be the delay target in slots. Between two consecutive transmission attempts of the \emph{same} packet, the earliest the transmitter can react, i.e., learn the outcome and schedule the next attempt when needed, is after an effective round-trip time
\begin{equation}
    \mathrm{RTT} \triangleq 1 + \ddecod + \dfeedb
\end{equation}
(in slots), where the ``$1$'' accounts for the transmission slot itself. 

If no blocked slot occurs during the service of the tagged packet, the maximum number of attempts that can be completed within deadline $\bar d$ is
\begin{equation}
M_{\bar d}^{\mathrm{noCS}}
=
\min\left(
M,
\left\lfloor
\frac{\bar d+\dfeedb}{\dfeedb+\ddecod+1}
\right\rfloor
\right).
\label{eq:secIR_kd}
\end{equation}
Here, the denominator represents the RTT, while the numerator captures the delay target after subtracting the feedback delay $\dfeedb$ for the last transmission attempt, since a successfully decoded packet can be forwarded to the upper layers without waiting for the corresponding feedback.
Accordingly, the service-\ac{DVP}, i.e., the packet is still not completed by $\bar d$ slots, is
\begin{equation}
\mathbb{P}(\dserv^{\mathrm{noCS}}>\bar d)
=
\prod_{i=1}^{M_{\bar d}^{\mathrm{noCS}}} p_i .
\label{eq:secIR_servDelay}
\end{equation}

If a blocked slot occurs during the service period, at least one transmission opportunity is lost, effectively reducing the delay target by one slot in our bound. The corresponding attempt budget becomes
\begin{equation}
M_{\bar d}^{\mathrm{CS}}
=
\min\left(
M,
\left\lfloor
\frac{\bar d+\dfeedb-1}{\dfeedb+\ddecod+1}
\right\rfloor
\right),
\label{eq:secIR_kd_xi}
\end{equation}
and
\begin{equation}
\mathbb{P}(\dserv^{\mathrm{CS}}>\bar d)
=
\prod_{i=1}^{M_{\bar d}^{\mathrm{CS}}} p_i .
\label{eq:secIR_servDelay_xi}
\end{equation}

Assume $\dwait=k$, i.e., the packet has already spent $k$ slots waiting. The residual delay target for service is then $\bar d-k$. Replacing $\bar d$ by $\bar d-k$ in \eqref{eq:secIR_kd}-\eqref{eq:secIR_kd_xi} yields
\begin{align}
M_{\bar d-k}^{\mathrm{noCS}}
&=
\min\left(
M,
\left\lfloor
\frac{\bar d-k+\dfeedb}{\dfeedb+\ddecod+1}
\right\rfloor
\right),
\label{eq:secIR_kd_residual}
\\
M_{\bar d-k}^{\mathrm{CS}}
&=
\min\left(
M,
\left\lfloor
\frac{\bar d-k+\dfeedb-1}{\dfeedb+\ddecod+1}
\right\rfloor
\right).
\label{eq:secIR_kd_residual_xi}
\end{align}

Over at most $M_{\bar d-k}^{\mathrm{noCS}}$ attempts, the probability that the service interval is affected by at least one blocked slot is upper-bounded by
\begin{equation}
\alpha_{\bar d-k}
=
\min\left\{1,\frac{M_{\bar d-k}^{\mathrm{noCS}}}{\xi}\right\}.
\label{eq:upperbound_probability}
\end{equation}

Using the two service cases (no blocked slot versus at least one blocked slot) and weighting them by $\alpha_{\bar d-k}$ yields the service-\ac{DVP} and the total \ac{DVP} in Lemma~\ref{lemma_ir_service_bound_cso} and Theorem~\ref{theorem_ir_dvp_cso}, respectively. 
The proofs are also provided in Appendices~\ref{proof_lemma_service} and \ref{proof_theorem_dvp}.

\begin{lemma}\label{lemma_ir_service_bound_cso}
For a given realized waiting delay $k$ such that $0\le k\le \lfloor \bar d \rfloor$, the conditional service-delay violation probability satisfies
\begin{align}
\mathbb{P}(\dserv>\bar d-k \mid \dwait=k)
\le\;&\\
\left(1-\alpha_{\bar d-k}\right)\prod_{i=1}^{M_{\bar d-k}^{\mathrm{noCS}}} p_i
&  + \alpha_{\bar d-k}\prod_{i=1}^{M_{\bar d-k}^{\mathrm{CS}}} p_i.
\nonumber
\end{align}
\end{lemma}

Combining Lemma~\ref{lemma_ir_service_bound_cso} with the waiting-delay \ac{PMF} in \eqref{eq:secIR_waitDelay} gives an upper bound on the total \ac{DVP}.

\begin{theorem}\label{theorem_ir_dvp_cso}
Let $\bar d=d/T$. The total delay violation probability satisfies
\begin{align}
\mathbb{P}&(\dtot>\bar d)
\le\;\\
&\sum_{k = 0}^{\lfloor \bar d \rfloor} f_{\dwait}(k)\!
\Big[
\left(1-\alpha_{\bar d-k}\right)\prod_{i=1}^{M_{\bar d-k}^{\mathrm{noCS}}} p_i + \alpha_{\bar d-k}\prod_{i=1}^{M_{\bar d-k}^{\mathrm{CS}}} p_i \Big].
\nonumber
\end{align}
\end{theorem}

\section{Numerical Evaluation}\label{sec:simulations}

In this section, we evaluate the performance of the proposed model in estimating the upper bound for DVP through a two-stage evaluation. We first verify that the proposed model provides an upper bound for the DVP, using a numerical evaluation implemented in \texttt{MATLAB}, under the system model assumptions introduced in the earlier sections. This validation focuses on the modeled queueing dynamics rather than the detailed 5G protocol stack. 
We then move to a more realistic 5G environment using the packet-level \texttt{ns-3} 5G-LENA simulator, which captures practical effects not explicitly modeled in the \texttt{MATLAB} analysis, including a detailed channel model and protocol stacks. The 5G-LENA simulations use an end-to-end NR stack with time-slotted scheduling, \ac{HARQ} operation, realistic protocol timing, and a 3GPP indoor channel model. 
In the \texttt{ns-3} 5G-LENA simulator, packet errors are determined by the link abstraction based on the instantaneous \ac{SNR} and \ac{MCS} index. 
Moreover, in this work, for each configuration point, the reference \ac{DVP} is measured from more than $10^6$ packet transmissions.
Additionally, each (re)transmission uses a fixed allocation of $\nrb$ \acp{PRB} per slot, and for a given packet size $n$, we select the lowest \ac{MCS} index whose spectral efficiency $\eta$ satisfies $134.16\,\nrb\,\eta \ge n$. For both \texttt{MATLAB} analysis and \texttt{ns-3} 5G-LENA simulations, unless otherwise specified, the default values and ranges for the parameters are summarized in Table~\ref{table:params}.

To demonstrate the performance of our proposed model in estimating the DVP of \textit{5G Simulation} using 5G-LENA in Subsection\ref{subsec:results_discussion}, we have compared it with the state-of-the-art models, including:
\begin{enumerate}
    \item \textit{Max Throughput}~\cite{Vishnu2025}: A variant of the \textit{Proposed} model that does not account for periodic \ac{CS} blockage during service, i.e., $\alpha_{\bar d-k}=0$. By assuming that no transmission opportunities are lost due to \ac{CSP}, this model represents a maximum-throughput case with the highest service opportunity, resulting in an optimistic, lower estimate of the \ac{DVP} bound. This benchmark isolates the impact of \ac{CSP} on the \ac{DVP} bound.
    \item \textit{Fixed Tx Rate}: A fixed-transmission-rate queueing baseline inspired by~\cite{FixedTx}, parameterized using the \ac{PER} values obtained from the 5G-LENA link-level results. Since this baseline is \ac{ARQ}-based, it uses only $p_1$ for retransmissions.
\end{enumerate}

To facilitate reproducibility, the simulation code with detailed parameter settings is made publicly available\footnote{Available at \url{https://github.com/wwtkddnjsww/Dvp5GHarq}.}. Additional results for larger delay targets are provided in Appendix~\ref{app:AdditionalResults}.

\begin{table}
\setlength{\belowcaptionskip}{6pt}
\centering
\caption{Default values and range of key parameters.}
\label{table:params}
\begin{tabular}{|l|l|l|}
\hline
Parameter             & Default value     & Range                                                                                                                            \\ \hline
$T$                   & 1 ms              & - \\
$n$                   & 100$\times$8 bits & -                                                                                                  \\
$\eta_{\mathrm{min}}$ & 0.2344         & -                                                                                                                                \\
$\eta_{\mathrm{max}}$ & 5.5547         & -                                                                                                                                \\
$\nrb$                &  -             & $\left\{\left\lceil\frac{n}{180\eta_{\mathrm{max}}}\right\rceil,\left\lceil\frac{n}{180\eta_{\mathrm{min}}}\right\rceil\right\}$ \\
$\snr$                & 10 dB          & -                                                                                                              \\
$\ddecod$             & 0.5 ms (1/2 slots)        & -                                                                                                                                \\
$\dfeedb$              & 0.5 ms (1/2 slots)       & -                                                                                                                                \\
$d$                   & 5 ms ($\bar d =$ 5 slots)           & $[2, 10]$ ms                                                                                                                     \\
$f$                   & $\frac{1}{3}$  & -                   \\ 
$\xi$                 & 80             & {80, 320} \\
$\qmax$ & 16 & - \\\hline
\end{tabular}
\end{table}

\begin{figure}[t]
\centering
\subfloat[Effect of arrival rate $f$ and delay target $d$.\label{fig:val_arr}]{
    \includegraphics[width=0.8\linewidth]{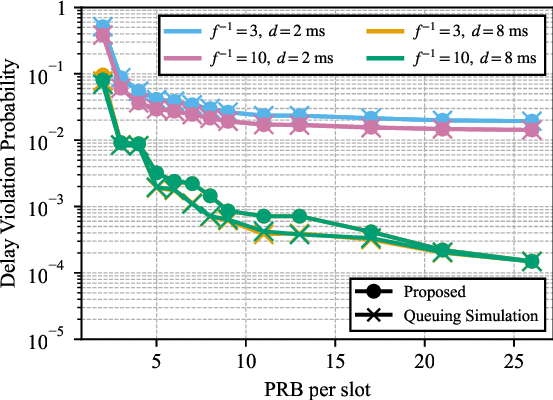}
}
\vspace{1mm}
\subfloat[Effect of SNR $\gamma$ and delay target $d$.\label{fig:val_snr}]{
    \includegraphics[width=0.8\linewidth]{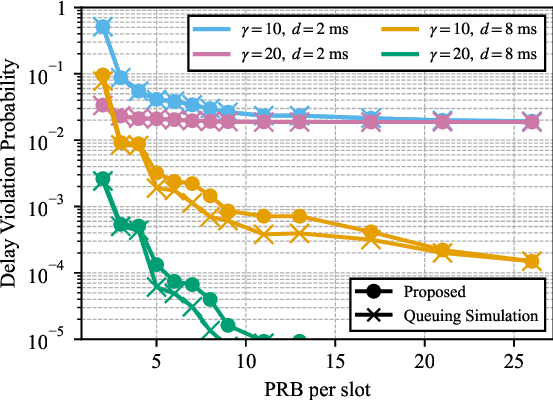}
}
\caption{Validation of the \textit{Proposed} model against queueing simulation.}
\label{fig:val}
\end{figure}

\subsection{Validation of the Proposed Model}\label{subsec:val}
We first validate whether the \textit{Proposed} model (i.e., \ac{DVP} upper bound in Theorem~\ref{theorem_ir_dvp_cso}) serves as an upper bound on the \textit{Queueing Simulation} implemented in \texttt{MATLAB}, under the assumptions of the system model.
Figure~\ref{fig:val} presents this comparison under two representative system settings. In Figure~\ref{fig:val_arr}, $f=1/3$ and $f=1/10$ are compared with fixed $\gamma=10$ dB and $\xi=80$. In Figure~\ref{fig:val_snr}, $\gamma=10$ dB and $\gamma=20$ dB are compared with fixed $f=1/3$ and $\xi=80$.
In both cases, two representative delay targets, $d=2$ ms and $d=8$ ms, are considered to assess the accuracy of the model under stringent and relaxed delay constraints.
In both Figures~\ref{fig:val_arr} and~\ref{fig:val_snr}, the \textit{Proposed} model remains identical to or slightly above the simulation results over the $\nrb$ range. 
For the more stringent delay target $d=2$, the two results almost overlap, indicating that the \textit{Proposed} model provides a very tight upper bound while characterizing both the sharp DVP decrease in the low-$\nrb$ regime and the floor behavior in the high-$\nrb$ regime. 
For the more relaxed delay target $d=8$ ms, the result of the \textit{Proposed} model becomes slightly higher than the queueing simulation result, thereby providing a close upper bound.

\subsection{Performance Comparison and Impact Analysis}\label{subsec:results_discussion}

\begin{figure}[t]
\centering
\subfloat[$d=2$ ms.\label{fig:arr2ms}]{
    \includegraphics[width=0.8\linewidth]{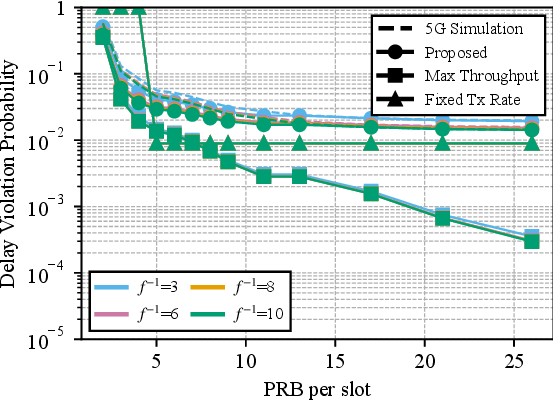}
}
\vspace{1mm}
\subfloat[$d=5$ ms.\label{fig:arr5ms}]{
    \includegraphics[width=0.8\linewidth]{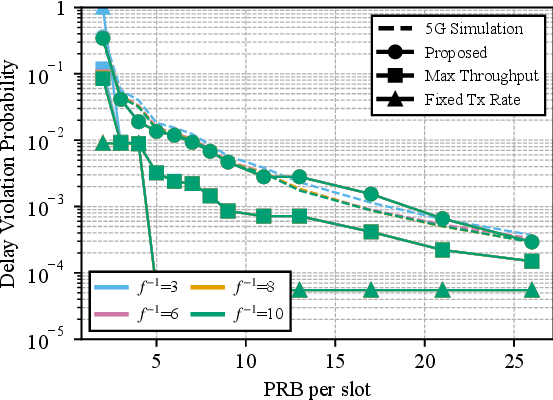}
}
\caption{DVP vs. allocated $\nrb$ per slot for different arrival rates $f$.}
\label{fig:arr}
\end{figure}

\subsubsection{Impact of the Offered Load (Arrival Rate $f$)}\label{subsubsec:results_arrival}
Figure~\ref{fig:arr} illustrates the \ac{DVP} as a function of $\nrb$ for various arrival probabilities $f$, considering both $d=2$~ms and $d=5$~ms delay targets. Two key trends emerge from these results. First, under the stringent delay target of $d=2$~ms, shown in Figure~\ref{fig:arr2ms}, the \ac{DVP} exhibits a noticeable \emph{floor} at moderate to large values of $\nrb$, indicating a delay-limited regime where increasing radio resources alone cannot fully prevent delay violations. In this regime, a higher arrival rate $f$ leads to a greater likelihood of queue build-up, which in turn causes a slight increase in the \ac{DVP}.
Second, when the delay target is relaxed to $d=5$~ms, illustrated in Figure~\ref{fig:arr5ms}, the \ac{DVP} becomes significantly less sensitive to changes in $f$ across the considered range of $\nrb$. This suggests that the additional time budget is sufficient to accommodate moderate queue backlogs and occasional retransmissions.

Across both delay target scenarios, the \textit{Proposed} bound closely follows the results obtained from the 5G-LENA simulation, while the \textit{Max Throughput} benchmark becomes overly optimistic at larger $\nrb$ values, reflecting its omission of periodic lost transmission opportunities. The \textit{Fixed Tx Rate} baseline captures the general trend but differs in both level and slope, highlighting the limitations of a sequential fixed-rate abstraction in the presence of \ac{MCS} adaptation and \ac{HARQ}-driven timing.

\begin{figure}[!t]
\centering
\subfloat[$d=2$ ms.\label{fig:snr2ms}]{
    \includegraphics[width=0.8\linewidth]{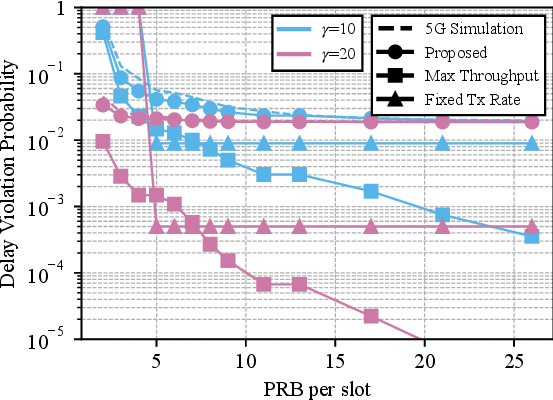}
}
\vspace{1mm}
\subfloat[$d=5$ ms.\label{fig:snr5ms}]{
    \includegraphics[width=0.8\linewidth]{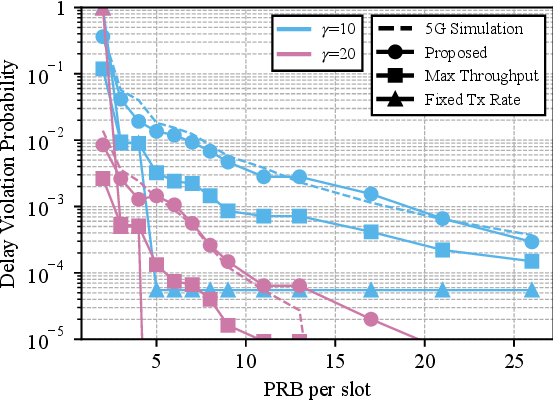}
}
\vspace{1mm}
\subfloat[$d=8$ ms.\label{fig:snr8ms}]{
    \includegraphics[width=0.8\linewidth]{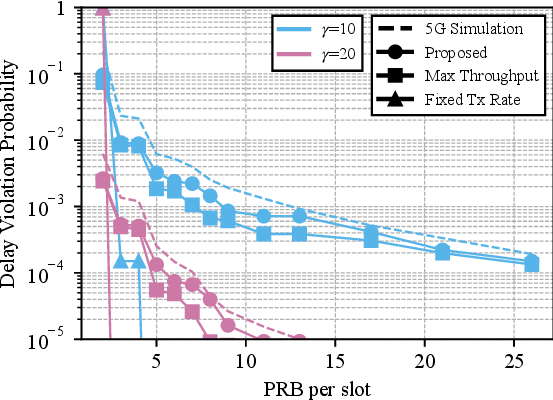}
}
\caption{DVP vs. allocated $\nrb$ per slot for different SNR $\gamma$.}
\label{fig:snr}
\end{figure}

\subsubsection{Impact of Link Quality (\ac{SNR} $\gamma$)}\label{subsubsec:results_snr}
Figure~\ref{fig:snr} shows the \ac{DVP} as a function of $\nrb$ for $\gamma \in \{10, 20\}$~dB across three different delay targets.
The results reveal a clear transition from a \emph{reliability-limited} regime to a \emph{delay-limited} regime. At lower values of $\nrb$, increasing the \ac{SNR} $\gamma$ leads to a substantial reduction in the \ac{DVP}, as improved physical layer reliability decreases the need for retransmissions and shortens service times. However, as $\nrb$ increases, particularly for the stringent delay target of $d=2$~ms, shown in Figure~\ref{fig:snr2ms}, the \ac{DVP} curves approach a floor. In this region, retransmissions become infrequent, and the remaining delay violations are primarily driven by tight end-to-end timing constraints, including queueing and protocol-related delays. As a result, further improvements in \ac{SNR} yield diminishing returns.

The \textit{Proposed} model accurately captures this transition and remains closely aligned with the 5G simulation results across all three delay targets. In contrast, the \textit{Max Throughput} benchmark increasingly underestimates the \ac{DVP} at higher $\nrb$ values, especially at high \ac{SNR}, where performance becomes more sensitive to timing effects, including periodic \ac{CS} blockage, than to packet errors. The \textit{Fixed Tx Rate} baseline also tends to be optimistic in several regimes, reflecting the limitations of its simplified service abstraction.

\begin{figure}[!t]
\centering
\subfloat[$d=2$ ms.\label{fig:cso2ms}]{
    \includegraphics[width=0.8\linewidth]{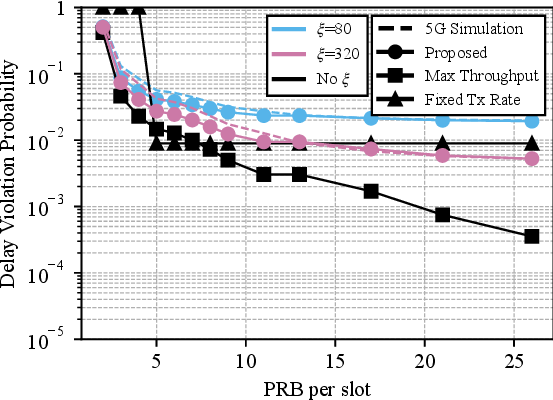}
}
\vspace{1mm}
\subfloat[$d=5$ ms.\label{fig:cso5ms}]{
    \includegraphics[width=0.8\linewidth]{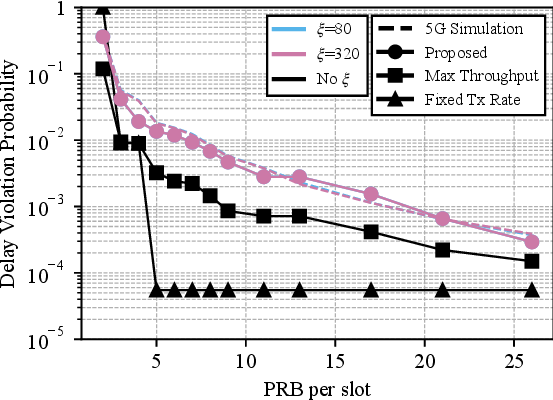}
}
\vspace{1mm}
\subfloat[$d=8$ ms.\label{fig:cso8ms}]{
    \includegraphics[width=0.8\linewidth]{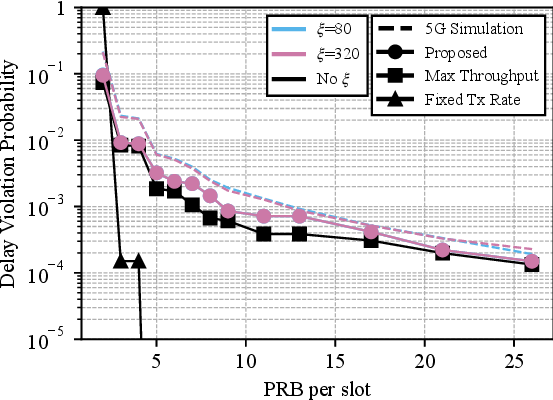}
}
\caption{DVP vs. allocated $\nrb$ per slot for different CSP $\xi$.}
\label{fig:cso}
\end{figure}

\subsubsection{Impact of Periodic \ac{CS} (\ac{CSP} $\xi$)}\label{subsubsec:results_csp}
Figure~\ref{fig:cso} illustrates the impact of \ac{CSP} by comparing scenarios with $\xi=80$ and $\xi=320$. A larger value of $\xi$ corresponds to less frequent blocked slots, thereby increasing the availability of transmission opportunities for data.
The effect of \ac{CSP} is most pronounced under the stringent delay target of $d=2$~ms, shown in Figure~\ref{fig:cso2ms}. In this case, reducing the frequency of blockage (i.e., increasing $\xi$) leads to a noticeable decrease in the \ac{DVP}, highlighting that even occasional loss of transmission opportunities can be critical when the latency budget is tight. As the delay target becomes more relaxed (for example, $d=8$~ms, shown in Figure~\ref{fig:cso8ms}, the influence of \ac{CSP} diminishes significantly, since the system can accommodate more transmission attempts and absorb occasional blocked slots within the larger time budget.

The \textit{Proposed} model closely matches the 5G-LENA simulation results for both \ac{CSP} settings, demonstrating that explicitly modeling \ac{CSP} in both queue evolution and service feasibility is essential for accurate \ac{DVP} prediction. In contrast, the \textit{Max Throughput} benchmark consistently underestimates the \ac{DVP}, as it assumes uninterrupted data-slot availability, while the \textit{Fixed Tx Rate} baseline does not capture the \ac{CSP}-dependent behavior of the \ac{HARQ} pipeline.

In a nutshell, these numerical results support three main conclusions. First, under stringent delay targets, the \ac{DVP} often exhibits a \emph{floor} that is dominated by protocol timing and queueing effects, rather than by physical layer reliability alone. Second, improvements in \ac{SNR} or resource allocation primarily reduce the \ac{DVP} in the reliability-limited regime; once the system becomes delay-limited, timing constraints and slot availability become the dominant factors. Third, periodic \ac{CS} blockage can have a significant impact under tight latency budgets, and neglecting this effect leads to overly optimistic \ac{DVP} predictions.

\section{Conclusion}\label{sec:conclusion}
This paper introduced a realistic \ac{DVP} characterization for slot-based 5G systems with \ac{HARQ}. Unlike simplified models, our framework explicitly accounts for key delay components, including queueing, decoding and processing, feedback, and transmission opportunities lost to \ac{CSP}, and reflects parallel \ac{HARQ} operation. We decomposed end-to-end delay into waiting and service components, derived the waiting-delay distribution from stationary system behavior, and combined it with a timing-consistent service-delay analysis to obtain a tractable upper bound on the \ac{DVP}.
Using per-attempt \ac{PER} data from \texttt{ns-3} 5G-LENA simulations, our results show that the proposed model closely matches packet-level simulation across a wide range of scenarios. The findings highlight that neglecting decoding/feedback timing or \ac{CSP}-induced blockage can lead to overly optimistic \ac{DVP} estimates, especially under tight delay targets where even minor queueing or lost transmission opportunities can dominate tail-delay performance. While not fully closed-form, our approach offers a tractable tool for \ac{DVP}-centric evaluation of 5G \ac{HARQ} systems, supporting informed resource allocation and \ac{QoS}-aware design under realistic protocol timing.

\bibliographystyle{IEEEtran}
\bibliography{refs}

\begin{figure*}[!t]
\centering
\subfloat[$d=3$ ms.\label{fig:arr3ms}]{
    \includegraphics[width=0.32\textwidth]{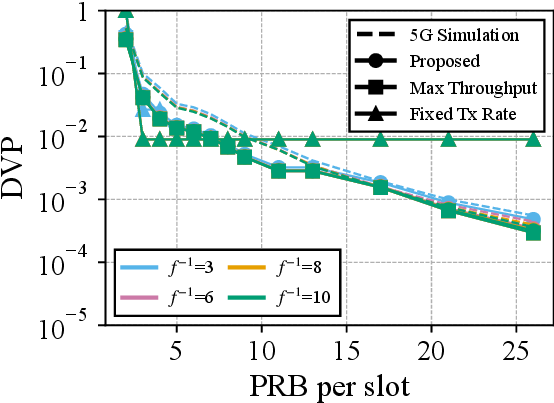}
}
\subfloat[$d=8$ ms.\label{fig:arr8ms}]{
    \includegraphics[width=0.32\textwidth]{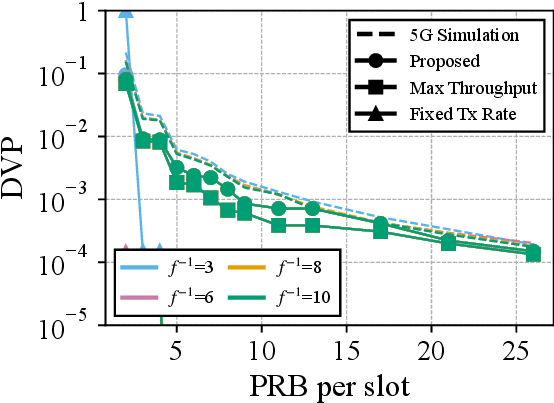}
}
\subfloat[$d=10$ ms.\label{fig:arr10ms}]{
    \includegraphics[width=0.32\textwidth]{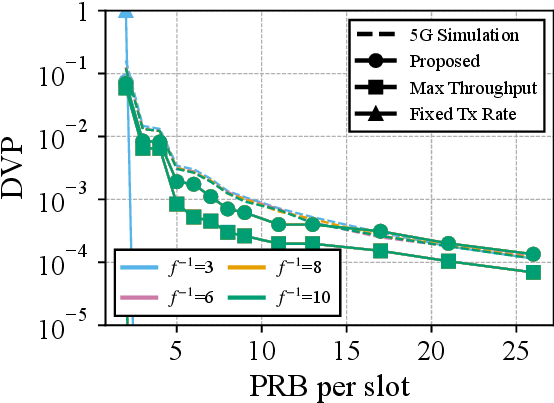}
}
\caption{DVP vs. allocated $\nrb$ per slot for different arrival rates $f$.}
\label{fig:arrapp}
\end{figure*}

\begin{figure*}[ht]
\centering
\subfloat[$d=3$ ms.\label{fig:snr3ms}]{
    \includegraphics[width=0.34\linewidth]{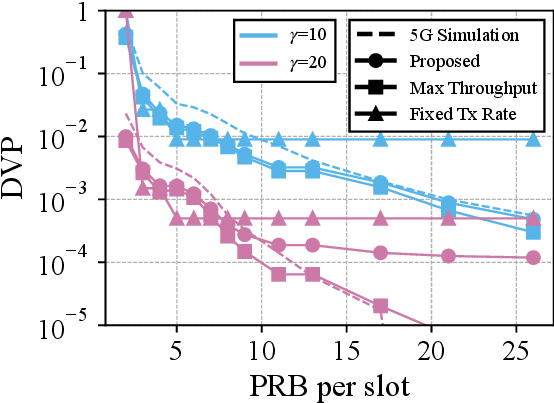}
} 
\subfloat[$d=10$ ms.\label{fig:snr10ms}]{
    \includegraphics[width=0.34\linewidth]{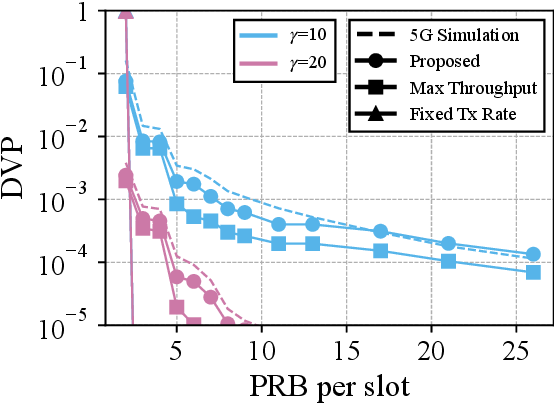}
}\\
\subfloat[$d=3$ ms.\label{fig:cso3ms}]{
    \includegraphics[width=0.34\linewidth]{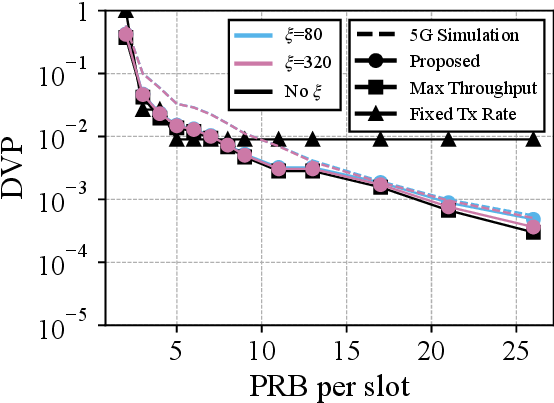}
} 
\subfloat[$d=10$ ms.\label{fig:cso10ms}]{
    \includegraphics[width=0.34\linewidth]{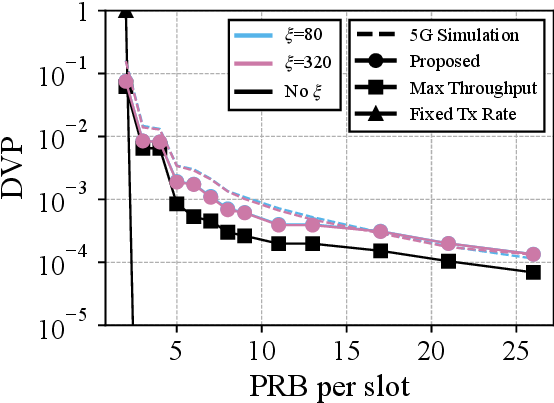}
}
\caption{DVP vs. allocated $\nrb$ per slot for different parameters: (a) and (b) SNR $\gamma$, (c) and (d) different CSP $\xi$.}
\label{fig:snrapp}
\end{figure*}

\appendices
\section{Proof of Lemma~\ref{lemma_ir_service_bound_cso}}
\label{proof_lemma_service}
Let $\Xi$ denote the event that at least one HARQ transmission opportunity within the service period is blocked by CS, and $\Xi^c$ denote its complement. Then,
\begin{align}
\mathbb{P}(\dserv\!>\!\bar d-k\!\mid\!\dwait\!=\!k)&\!=\!\mathbb{P}(\Xi^c\!\mid\!\dwait\!=\!k)\ \!\mathbb{P}(\dserv^{\mathrm{noCS}}\!>\!\bar d\!-\!k)\nonumber\\
&~~\!+\!\mathbb{P}(\Xi\!\mid\!\dwait\!=\!k)\ \!\mathbb{P}(\dserv^{\mathrm{CS}}\!>\!\bar d\!-\!k).
\label{eq:proof_event_split}
\end{align}
Let
\begin{align}
A_k
&\triangleq
\mathbb{P}(\dserv^{\mathrm{noCS}}>\bar d-k),\nonumber
\\
B_k
&\triangleq
\mathbb{P}(\dserv^{\mathrm{CS}}>\bar d-k).\nonumber
\end{align}
Then \eqref{eq:proof_event_split} can be rewritten as
\begin{align}
\mathbb{P}(\dserv>\bar d-k\!\mid\!\dwait\!=\!k)\!&=\!A_k\!+\!\mathbb{P}(\Xi\!\mid\!\dwait\!=\!k)(B_k\!-\!A_k),
\label{eq:proof_event_split_rewrite}
\end{align}
where we used $\mathbb{P}(\Xi^c \!\mid\! \dwait\!=\!k) \!=\! 1\!-\!\mathbb{P}(\Xi \!\mid\! \dwait\!=\!k)$.
Now let $N$ be the actual number of HARQ transmission opportunities utilized by the packet. By construction,
\begin{equation}
N \le \max\left(M_{\bar d-k}^{\mathrm{CS}},M_{\bar d-k}^{\mathrm{noCS}}\right)\leq M_{\bar d-k}^{\mathrm{noCS}}.
\end{equation}
As one out of every $\xi$ slots is reserved for CS, we have
\begin{equation}
\mathbb{P}(\Xi \mid \dwait=k)
\le
\frac{N}{\xi}
\le
\frac{M_{\bar d-k}^{\mathrm{noCS}}}{\xi}
=
\alpha_{\bar d-k}.
\label{eq:proof_cso_bound}
\end{equation}
Applying \eqref{eq:proof_cso_bound} into \eqref{eq:proof_event_split_rewrite} yields
\begin{align}
\mathbb{P}(\dserv\!>\!\bar d&\!-\!k\!\mid\!\dwait\!=\!k)\!\le\!A_k\!+\!\alpha_{\bar d-k}(B_k\!-\!A_k)\!=\!
\nonumber\\
&\left(1\!-\!\alpha_{\bar d-k}\right)\!\prod_{i=1}^{M_{\bar d-k}^{\mathrm{noCS}}}\!p_i\!+\!\alpha_{\bar d-k}\!\prod_{i=1}^{M_{\bar d-k}^{\mathrm{CS}}}\!p_i,\nonumber
\end{align}
thus completing the proof.
\hfill\qedsymbol

\section{Proof of Theorem~\ref{theorem_ir_dvp_cso}}
\label{proof_theorem_dvp}
Since $\dtot\!=\!\dwait+\dserv$, conditioning on the waiting delay yields
\begin{align}
\mathbb{P}(\dtot>\bar d)
&=
\sum_{k=0}^{\lfloor \bar d \rfloor}
\mathbb{P}(\dwait=k)\,
\mathbb{P}(\dserv>\bar d-k \mid \dwait=k).\nonumber
\end{align}
Applying Lemma~\ref{lemma_ir_service_bound_cso} and using $\mathbb{P}(\dwait=k)=f_{\dwait}(k)$ given in~\eqref{eq:secIR_waitDelay} gives
\begin{align}
\mathbb{P}&(\dtot\!>\!\bar d)\!\le\!\nonumber\\
&
\sum_{k=0}^{\lfloor \bar d \rfloor}\!f_{\dwait}(k)\!\left[
\!\left(1\!-\!\alpha_{\bar d-k}\right)\!\prod_{i=1}^{M_{\bar d-k}^{\mathrm{noCS}}}\!p_i\!+\!\alpha_{\bar d-k}\!\prod_{i=1}^{M_{\bar d-k}^{\mathrm{CS}}}\!p_i\!\right]\!,\nonumber
\end{align}
thus completing the proof.\hfill\qedsymbol

\section{Additional Numerical Evaluations}
\label{app:AdditionalResults}
To further assess the validity of the proposed model, this appendix presents additional numerical evaluations for delay targets not included in Section~\ref{sec:simulations}. Figures~\ref{fig:arrapp} and~\ref{fig:snrapp} show that the proposed model remains tight beyond the representative cases shown in the main text. 
They are included in the appendix for brevity, since the qualitative trends remain unchanged and the differences among the models, particularly between the \textit{Proposed} and \textit{Max Throughput} models, become less pronounced at these delay targets.

\balance

\end{document}

%% file: acronyms.tex
\begin{acronym}
  \acro{URLLC}{ultra-reliable low latency communication}
  \acro{HARQ}{hybrid automatic repeat request}
  \acro{ARQ}{automatic repeat request}
  \acro{eMBB}{enhanced mobile broadband}
  \acro{mMTC}{massive machine-type communications}
  \acro{QoS}{quality-of-service}
  \acro{PER}{packet error rate}
  \acro{CS}{control signaling}
  \acro{DVP}{delay violation probability}
  \acro{FIFO}{first-in first-out}
  \acro{IR}{incremental redundancy}
  \acro{AWGN}{additive white Gaussian noise}
  \acro{FBL}{finite-blocklength}
  \acro{RTT}{round-trip time}
  \acro{CSP}{control signal periodicity}
  \acro{SNR}{signal-to-noise ratio}
  \acro{UE}{user equipment}
  \acro{UL}{uplink}
  \acro{DL}{downlink}
  \acro{MCS}{modulation and coding scheme}
  \acro{RB}{resource block}
  \acro{ACK}{acknowledgment}
  \acro{NACK}{negative acknowledgment}
  \acro{gNB}{next-generation node B}
  \acro{i.i.d.}{independent and identically distributed}
  \acro{OFDM}{orthogonal frequency-division multiplexing}
  \acro{NR}{New Radio}
  \acro{DTMC}{discrete-time Markov chain}
  \acro{PMF}{probability mass function}
  \acro{PRB}{physical resource block}
  \acro{PER}{packet error rate}
\end{acronym}